\documentclass[conference]{IEEEtran}
\IEEEoverridecommandlockouts

\usepackage{xcolor}
\usepackage{subfigure}
\usepackage{color}
\usepackage{graphicx}
\usepackage{balance}
\usepackage{amsmath}
\usepackage{amsfonts}
\usepackage{amssymb}% http://ctan.org/pkg/amssymb
\usepackage{pifont}% http://ctan.org/pkg/pifont
\usepackage{times}
\usepackage{paralist}
\usepackage{textcomp}
\usepackage[hyphens]{url}
\usepackage[font={footnotesize}]{caption}
\usepackage{units}
\usepackage{enumitem}
\usepackage{amsmath,colortbl}
\usepackage{cite,comment}
\usepackage[nolist]{acronym}
\usepackage{soul}
\usepackage{array}
\usepackage{tabularx}
\usepackage{multirow}
\usepackage{lipsum}
\begin{acronym}[ACRONYM]
  \acro{3GPP}{the third generation partnership project}
  \acro{4G}{fourth generation}
\acro{5G}{fifth generation}
\acro{6G}{sixth generation}
\acro{AI}{artificial intelligence}
\acro{AoA}{angle-of-arrival}
\acro{AoD}{angle-of-departure}
\acro{BS}{base station}
\acro{CDF}{cumulative density function}
\acro{MIMO}{multi-input multi-output}
\acro{IoT}{internet of things}
\acro{RSSI}{received signal strength indication}
\acro{CSI}{channel state information}
\acro{SV}{Saleh-Valenzuela}
\acro{ULA}{uniform linear array}
\acro{BS}{base stations}
\acro{CNN}{convolution neural network}
\acro{MFM}{Matlab for MIMO}
\acro{CDF}{cumulative distribution function}
\acro{RTI}{radio tomographic imaging}
\acro{JCAS}{joint communication and sensing}
\acro{ISAC}{integrated sensing and communication}
\acro{OFDM}{orthogonal frequency division multiplexing}
\acro{LOS}{line of sight}
\acro{NLOS}{non-line of sight}
\acro{DDP}{delay-Doppler profile}
\acro{PDP}{power delay profile}
\acro{RCS}{radar cross section}
\acro{SNR}{signal to noise ratio}
\acro{SNRs}{signal to noise ratios}
\acro{CPI}{coherent processing interval}
\acro{FD}{full duplex}
\acro{3GPP} {third generation partnership project}
\acro{CNN}{convolutional neural network}
\acro{CFAR}{constant false alarm rate}
\acro{PDF}{probability distribution function}
\acro{QPSK}{quadrature phase shift keying}
\end{acronym}

           \newcommand{\mbf}[1]{\mathbf{#1}}          
\newcommand{\mb}[1]{\mbox{#1}}
\newcommand{\SUB}[2]{#1_{\scriptsize \mbox{#2}}}
\newcommand{\MS}[1]{\scriptsize \mbox{#1}}
\newcommand{\suchthat}{\, \mid \,} % nice "such that"

\begin{document}
\bstctlcite{IEEEexample:BSTcontrol}

\title{Bi-Static Sensing in OFDM Wireless Systems for Indoor Scenarios}
\date{\today}
\author{Vijaya~Yajnanarayana and  Philipp Geuer \\
Ericsson Research 
\thanks{This work was supported by the Federal Ministry of Education and Research Germany within the project “KOMSENS-6G”, grant no: 16KISK127.}
}
\maketitle

\begin{abstract}
  The \ac{6G} systems will likely employ \ac{OFDM} waveform for performing the  joint task of sensing and communication. In this paper, we design an \ac{OFDM} system for \ac{ISAC} and propose a novel approach for passive target detection in an indoor deployment using a data driven AI approach. The \ac{DDP} and \ac{PDP}   is used to train the  proposed  AI-based detector. We analyze the detection performance of the proposed methods under \ac{LOS} and \ac{NLOS} conditions for various training strategies. We show that the proposed method provides $10\,\mb{dB}$ performance improvement over the baseline for $80\%$ target detection under \ac{LOS} conditions and the performance drops by $10-20\,\mb{dB}$ for \ac{NLOS} depending on the usecase scenarios. 

\end{abstract}


%\vspace{-0.5in}
\section{Introduction}
\label{sec:intro}
The successive generations of wireless systems have seen increase in the carrier frequency and bandwidth. This trend will likely to continue in \ac{6G}. The high frequency operation  of \ac{6G} systems together with wide operating bandwidth enable high resolution sensing  \cite{yang20196g, wymeersch-2021-integ-commun, behravan-2022-introd}. Passive target sensing involves typical radar functions such as detecting passive targets (i.e. non connected objects) and  estimate its state parameters. In this paper, we focus only on the detection of passive target and estimation of its position in a cluttered indoor environment.  The sensing extension to the ubiquitous communication infrastructure will extend sensing coverage  and aid in several use cases such as intruder detection, energy optimization by controlling the \ac{IoT} devices, tracking equipments among others. Compared to sensing using vision based sensors such as cameras, passive sensing using communication infrastructure ensures privacy and security \cite{zhang2021enabling, de2021convergent}.

\begin{figure}%[t]
  \centering
  % \fbox
  {\includegraphics[width = 0.8\columnwidth]{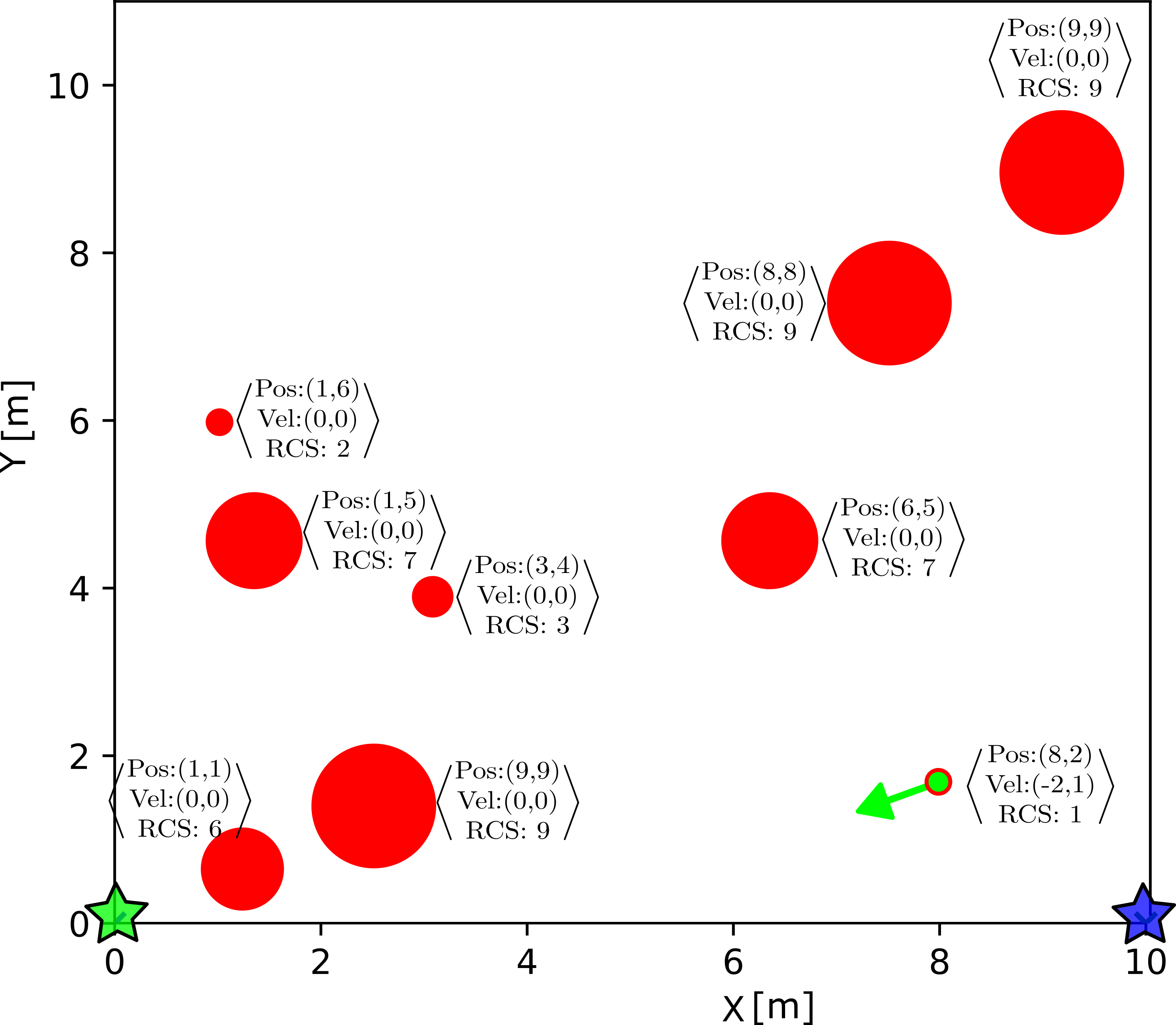}}
  \caption{Sensing a passive target of interest inside a room. Deployment consist of the transmitter and receiver on each corner of a room  as shown in green and blue stars respectively,  clutter is randomly placed and is shown in red circular patches. The passive target is shown with a green patch, with an  arrow showing the direction of the velocity.}
  \label{fig:deployment} \vspace{-3mm}
\end{figure}

\ac{OFDM} has been the de-facto waveform for communication since \ac{4G} and likely will continue to be used in \ac{6G} as it offers several benefits such as robustness against multipath fading, easy synchronization, flexibility in system design among others for communication. \ac{OFDM} is also good waveform for sensing, unlike pulsed or continuous waveform, \ac{OFDM} provides much needed frequency diversity there by improving sensing performance \cite{levanon2000multi, sturm2011waveform, sen2010adapt-ofdm}. In this paper, we exploit the \ac{OFDM} structure to multiplex sensing and  communication signals. The parameters for such an \ac{ISAC} signal is designed and  analyzed to ensure  sensing requirements for an indoor sensing scenario are met. We also propose several methods for passive target detection in indoor settings, a visualization of which is shown in the Fig.~\ref{fig:deployment}.

In a typical cellular infrastructure, a transmitting base station can exploit the reflections from the passive target using its receiver towards sensing, however this requires \ac{FD} operation. The \ac{FD} operation increases hardware complexity of the base station as it needs to handle challenges such as self interference, antenna coupling among others and thus multi/bi static sensing is considered \cite{behravan-2022-introd}. There exists several multi-static sensing works which exploit wireless signals [9-14], these works extract features like \ac{RSSI}, \ac{CSI} and micro-Doppler shifts in mid-band ($2-10\,\mb{GHz}$) communication signals towards radio sensing. Unlike in these works, in this paper, we propose to extract features such as  \ac{DDP} and \ac{PDP}  from high frequency mmWave signals towards sensing by employing AI methods.

In \cite{yajnanarayana2023eucnc}, high frequency mmWave band is considered for indoor sensing and in this work  stochastic geometric channel called Saleh-Valenzuela (SV) channel model is employed \cite{saleh-1987-statis-model}. High frequency channels tend to be environment specific and are hard to model using stochastic geometric channel models especially for indoor scenarios. In this paper, we employ a deterministic channel model and use \ac{DDP} processing at the receiver to extract relevant features to be fed into a novel AI architecture towards sensing. There also exist new ways of sensing which use \ac{RTI} for passive target sensing \cite{wilson-2010-radio-tomog}. These systems builds an attenuation image using many communication links for sensing and is not practical for many  indoor deployments.

In this paper, we propose methods for passive target detection using a joint communication and sensing waveform derived from \ac{OFDM}. The main contributions of this paper are summarized as follows:
\begin{itemize}
\item An hybrid method consisting of signal processing methods to extract  \ac{DDP} and \ac{PDP} profiles from  the received \ac{OFDM} signal and subsequently using them in AI algorithms towards target sensing.
\item Performance analysis of the proposed methods in terms of accuracy of target detection at different \ac{SNRs}. We analyze the detection performance under \ac{LOS} and \ac{NLOS} conditions when the target is  stationary or moving with a constant velocity.
\item We compare the performance of the proposed methods with minimum probability of error detector using energy detection. 
\end{itemize}

\begin{figure}%[t]
  \centering
  % \fbox
  {\includegraphics[width = 0.9\columnwidth]{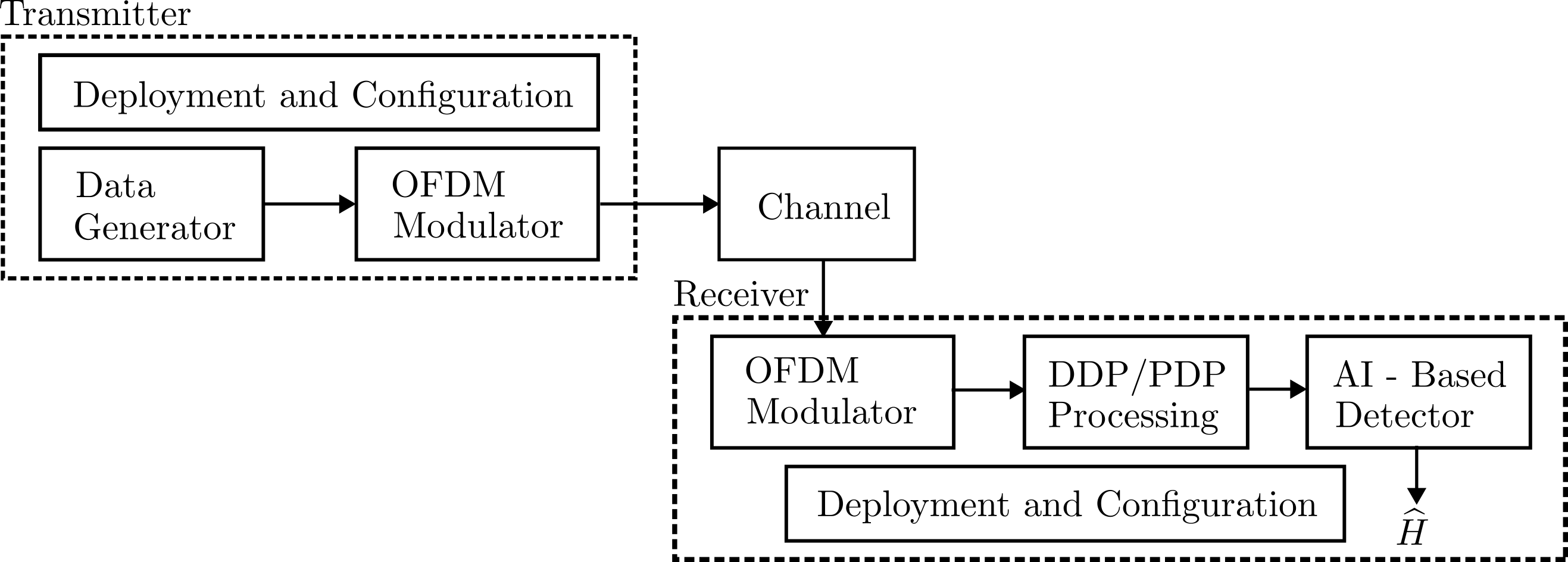}}
  \caption{The system model consist of the OFDM transmitter, which multiplexes the sensing reference signal with communication signal. At the receiver  \ac{DDP} or \ac{PDP} processing is performed which is subsequently exploited by AI detector towards sensing. }
  \label{fig:system_model} \vspace{-3mm}
\end{figure}

\begin{figure}%[t]
  \centering
  % \fbox
  {\includegraphics[width = 0.9\columnwidth]{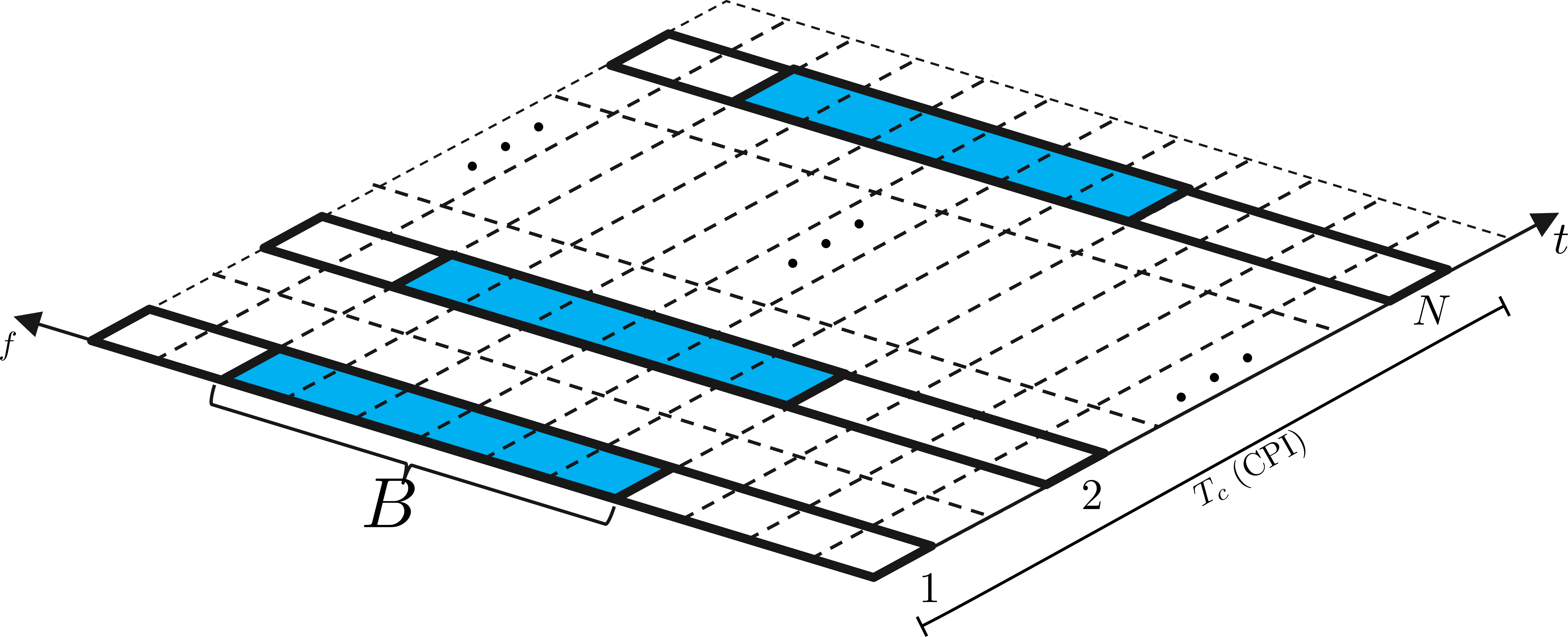}}
  \caption{OFDM Waveform for sensing a passive target inside a room. The sensing reference signal is periodically allocated shown in blue and is multiplexed with other communication signal. Duration for $N$ sensing OFDM pulses forms one \ac{CPI} which is used by the detector for sensing the target.}
  \label{fig:ofdm_waveform} \vspace{-3mm}
\end{figure}

\section{System Model}
\label{sec:sm}
In the following, we describe the system model for target sensing in the indoor environment. We consider an \ac{OFDM} system as shown in the Fig~\ref{fig:system_model}. The transmitter multiplexes the periodic sensing signals with communication information as shown in the Fig.~\ref{fig:ofdm_waveform}. We consider a known random \ac{QPSK} symbols as sensing payload while all zeros are used as communication payload. Sensing signals parameters such as periodicity, bandwidth and the number of OFDM symbols among others are designed to satisfy the indoor sensing requirements and are discussed in detail in the next section.  Such a multiplexed signal is impaired by the scatter point channel described in Section \ref{sec:cm}. At the receiver, depending on the use case, standard signal processing methods are applied to extract \ac{DDP} or \ac{PDP} features \cite{behravan-2022-introd} from the demodulated sensing signals. This is further processed by an AI detector towards inference on passive target detection.

\section{Waveform Design and System Parameterization}
A multiplexed \ac{OFDM} waveform consisting of sensing and communication as shown in Fig.~\ref{fig:ofdm_waveform} is employed. The sensing reference signal is periodically allocated to mimic an \ac{OFDM} radar. Typical indoor sensing problems require high Doppler resolution to resolve small target movements and high range resolution to separate a target from nearby clutter. The range resolution depends on the bandwidth, $B$, while the Doppler resolution depends on the \ac{CPI}, $T_c$ and  are given by $c/B$ and $c/(T_c f_c)$ for a bi-static system, respectively. Here $c$ and $f_c$ denotes speed of light and center frequency.  

We consider an OFDM numerology from \ac{3GPP} \ac{5G} system having a bandwidth, $B$ of $500\,\mb{MHz}$ and operating in mmWave bandwidth center frequency, $f_c$ of $28\,\mb{GHz}$. We consider $10\%$ of time-frequency resources are consumed for sensing, while 4-QAM is used. With such an OFDM numerology, it can be shown that a velocity and range resolution of $0.5\,\mb{m/s}$ and $0.6~\mb{m}$ respectively can be achieved by considering the \ac{CPI}, $T_c$ of around $20\,\mb{ms}$. The designed waveform parameters for the proposed indoor sensing scenario are shown in  Table~\ref{tab:params}.

\begin{table}%[t]
  \centering
  % \begin{tabular}{|p{1.5in}|p{1.5in}|}
  \begin{tabular}{|c|p{1.5in}|c|}
    \hline
    \textbf{Symbol} & \textbf{Quantity} & \textbf{Value} \\
    \hline
    $B$ & Bandwidth & $500\,\mb{MHz}$ \\
    $f_c$ & Center frequency & $28\,\mb{GHz}$\\
    $N_t$ & Number of Tx antenna & $1\mb{ (dipole antenna)}$\\
    $N_r$ & Number of Rx antenna & $1 \mb{ (dipole antenna)}$\\
    $M$ & Number of subcarriers & $1024$\\
    $\Delta_f$ & Subcarrier spacing & $480\,\mb{kHz}$\\
    $\mu$ & Percentage of resources allocated for sensing  &  $10\%$\\
    $N$ & Number of sensing symbols & $1024$\\
    $V_r$ & Velocity Resolution & $0.5\,\mb{m/s}$\\
    $R_r$ & Range Resolution & $0.6\,\mb{m}$\\
    $T_c$ & CPI & $20\,\mb{ms}$\\
    \hline                      
  \end{tabular}
  \caption{Waveform parameters}
  \label{tab:params}\vspace{-5mm}
\end{table}

\section{Channel model and deployment}
\label{sec:cm}

\subsection{LOS channel model}
\label{ss:los}
We consider $L$ clutter objects in the environment located at position $\mbf{p}_l\in \mathbb{R}^2$, $l=1,\ldots,L$  and a single passive target of interest at $\mbf{p}_t$, which is \ac{LOS} to transmitter and receiver. The stationary clutter is shown in red circular patches in the Fig.~\ref{fig:deployment} with its diameter indicating the size (denoted by \ac{RCS}) while the passive target is shown in the green patch having a non-zero velocity indicated by an arrow. A single transmitter and receiver is placed at two corners of the room. We consider a scatter point channel where clutter and target of interest are modeled as point scatterers \cite{chris2023eucnc}. The received signal is a sum of signals from a direct path and set of scattered paths from the target and clutter. The received  signal is given by

\begin{equation}
  \label{eq:rcv}
  r(t)=\displaystyle\sum_{l=0}^{L+1}b_ls(t-\tau_l)e^{j2\pi f_{D,l}t} + w(t),
\end{equation}
where $b_l$ denotes the gain of the $l\mb{-th}$ path. $l=0$ denotes the direct-path between transmitter and receiver while  $l=L+1$ denotes  the scattered path due to the target of interest. The scattered paths due to the $L$ clutter objects are indexed from $l=1,\ldots,L$. 
We consider stationary transmitter, receiver and clutter, therefore $f_{D,l}=0,\mb{ }l=0,\ldots,L$. $w(t)$ denotes the complex white Gaussian noise of variance $\beta^2$. The path gain of the direct path follows inverse square law with distance and is given by
\begin{equation}
  \label{eq:dir}
  b_0=\frac{G_tG_r \lambda_{c}^{2}}{4\pi||d_0||^2},
\end{equation}

where $d_0$ is the distance between transmitter and receiver, $G_t$ and $G_r$ are the antenna gain of the transmitter and receiver respectively, and $\lambda_c$ is the wavelength of the carrier. The scattered path gain due to the clutter and target follows the standard radar equation. During alternate ($\SUB{H}{1}$) hypothesis,
\begin{equation}
  \label{eq:radarequation}
  b_l=\frac{G_tG_r\sigma\lambda_{c}^{2}}{(4\pi)^3||d_{l,tx} ||^2||d_{l,rx} ||^2}, l=1,\ldots,L+1 , 
\end{equation}
where $\sigma$ is the \ac{RCS}, The $d_{l,tx}$, $d_{l,rx}$ are the distances to clutter or target from the transmitter and receiver respectively. During null hypothesis ($\SUB{H}{0}$), the clutter still follows the radar equation, but the target of interest does not exist, hence  we will have $b_{L+1}=0$ in \eqref{eq:radarequation}.

\subsection{NLOS channel model}
\label{ss:nlos}

\begin{figure}%[t]
  \centering
  % \fbox
  {\includegraphics[width = 0.9\columnwidth]{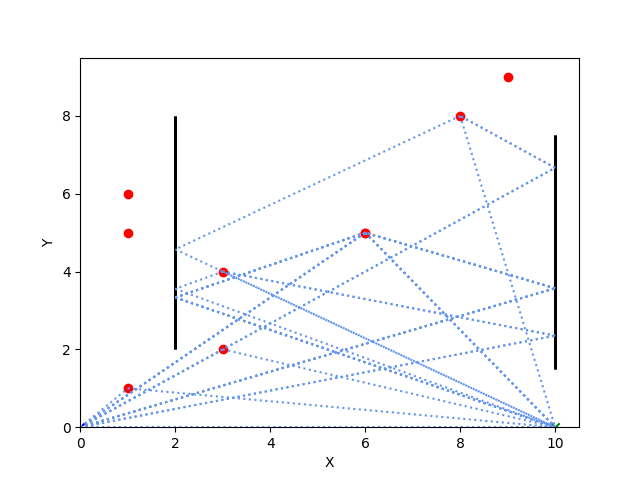}}
  \caption{NLOS deployment. Deployment consists of the same transmitter and receiver and clutter as the LOS scenario. The clutter also has the same RCS as in the LOS scenario, but for visibility the size is shown constant. In addition, two walls are added, with single specular reflections, and all paths are shown with blue dotted lines.}
  \label{fig:deployment-nlos} \vspace{-3mm}
\end{figure}

Below we extend the above discussed channel model towards \ac{NLOS}. The goal of this scenario is to examine the case, where the target of interest does not have a \ac{LOS} path to both transmitter and receiver. For this, the same \ac{LOS} deployment was used, but two walls were added as depicted in Fig.~\ref{fig:deployment-nlos}. Also, the same \ac{LOS} channel model was used, with the addition of specular reflections at the walls with a reflection loss of $3\,\mb{dB}$ with a maximum of one reflection between a scatterer and transmitter or receiver, respectively. The scattered paths between transmitter and receiver due to the stationary clutter are shown as blue dotted lines in Fig.~\ref{fig:deployment-nlos}.

\section{Methods}
\label{sec:methods}
\subsection{Feature Engineering and AI architecture}
\label{ss:arch}
We consider a hybrid approach for  target detection where key features are extracted from received signals using signal processing approach which are in turn used to train an AI detector. We consider two cases, one in which target of interest is moving, and another in which target of interest is stationary.
\subsubsection{Moving Target Case} Assuming omni-directional antennas for the transmitter and receiver, the received sensing reference symbol for the $k\mb{-th}$ sub-carrier and $n\mb{-th}$  symbol under \ac{LOS} condition is given by,
\begin{equation}
  \label{eq:dd}
  (\mbf{R})_{k,n}=\displaystyle\sum_{l=0}^{L+1}b_l(\mbf{S})_{k,n}e^{-j2\pi k\Delta_f\tau_l}e^{j2\pi nT\nu_l} + (\mbf{W})_{k,n}
\end{equation}
Where $\mbf{S}$ and $\mbf{R}$ are the transmit and receive OFDM symbols arranged in the form of matrix with sub-carrier and time in row and column dimension respectively. $\mbf{W}$ is  Gaussian matrix  representation of noise variance $\beta^2$.  Index $l$ should be interpreted similar to Section~\ref{sec:cm}.  $\tau_l$ and $\nu_l$ are the delay and Doppler for path $l$, and  $T$ is the \ac{OFDM} symbol duration.  From \eqref{eq:dd} we can see that the moving target or clutter will create row-wise and column-wise oscillations, whose frequencies depends on the delay and Doppler. Therefore, for the channel described, we can represent the received signal efficiently in delay-doppler domain. In our deployment, we assume a single moving target of interest among several stationary clutter  as depicted in Fig.~\ref{fig:deployment}, leading to $\nu_l=0, l\in\{0,1,\ldots,L\}$ and $\nu_{L+1}\ne 0$.  We first bring the received signal into delay-Doppler domain using Fourier processing described in \cite{behravan-2022-introd}, such a delay-Doppler frame is fed into a trained AI agent for target detection. A typical delay-Doppler frame for low and high Doppler scenarios are shown in the Fig.~\ref{fig:dd_pdp}(a) and Fig.~\ref{fig:dd_pdp}(b) respectively.
\subsubsection{Stationary Target Case}
\begin{figure}%[t]
  \centering
  % \fbox
  {\includegraphics[width = 0.95\columnwidth]{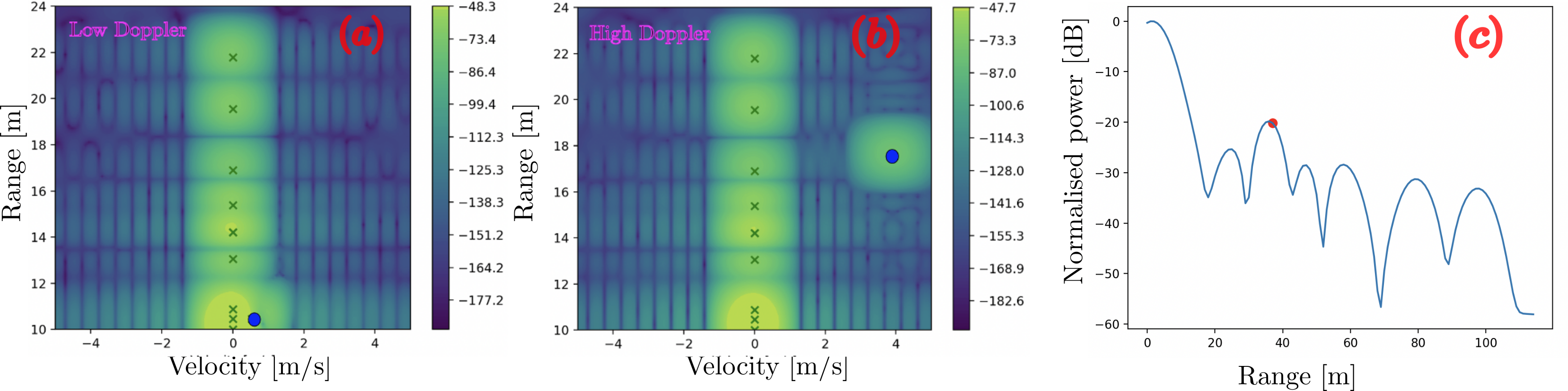}}
  \caption{Typical \ac{DDP} and \ac{PDP} profiles used by AI methods for target detection. The solid dot indicates the passive target's location in \ac{DDP} (velocity, range) and \ac{PDP} (range) domains.}
  \label{fig:dd_pdp} \vspace{-3mm}
\end{figure}

When the target of interest is stationary, then all the Doppler velocities including that of the target is zero, i.e., $\nu_l=0, l\in\{0,1,\ldots,L+1\}$. Therefore, $\eqref{eq:dd}$ reduces to
\begin{equation}
  \label{eq:pdp}
  (\mbf{R})_{k,n}=\displaystyle\sum_{l=0}^{L+1}b_l(\mbf{S})_{k,n}e^{-j2\pi k\Delta_f\tau_l} + (\mbf{W})_{k,n}.
\end{equation}
We compute \ac{PDP}  from $\eqref{eq:pdp}$  and feed it into a trained AI agent for target detection. A typical PDP for a stationary target is shown in the Fig.~\ref{fig:dd_pdp}(c). 
\begin{figure}
  \centering
  {\includegraphics[width = 0.8\columnwidth]{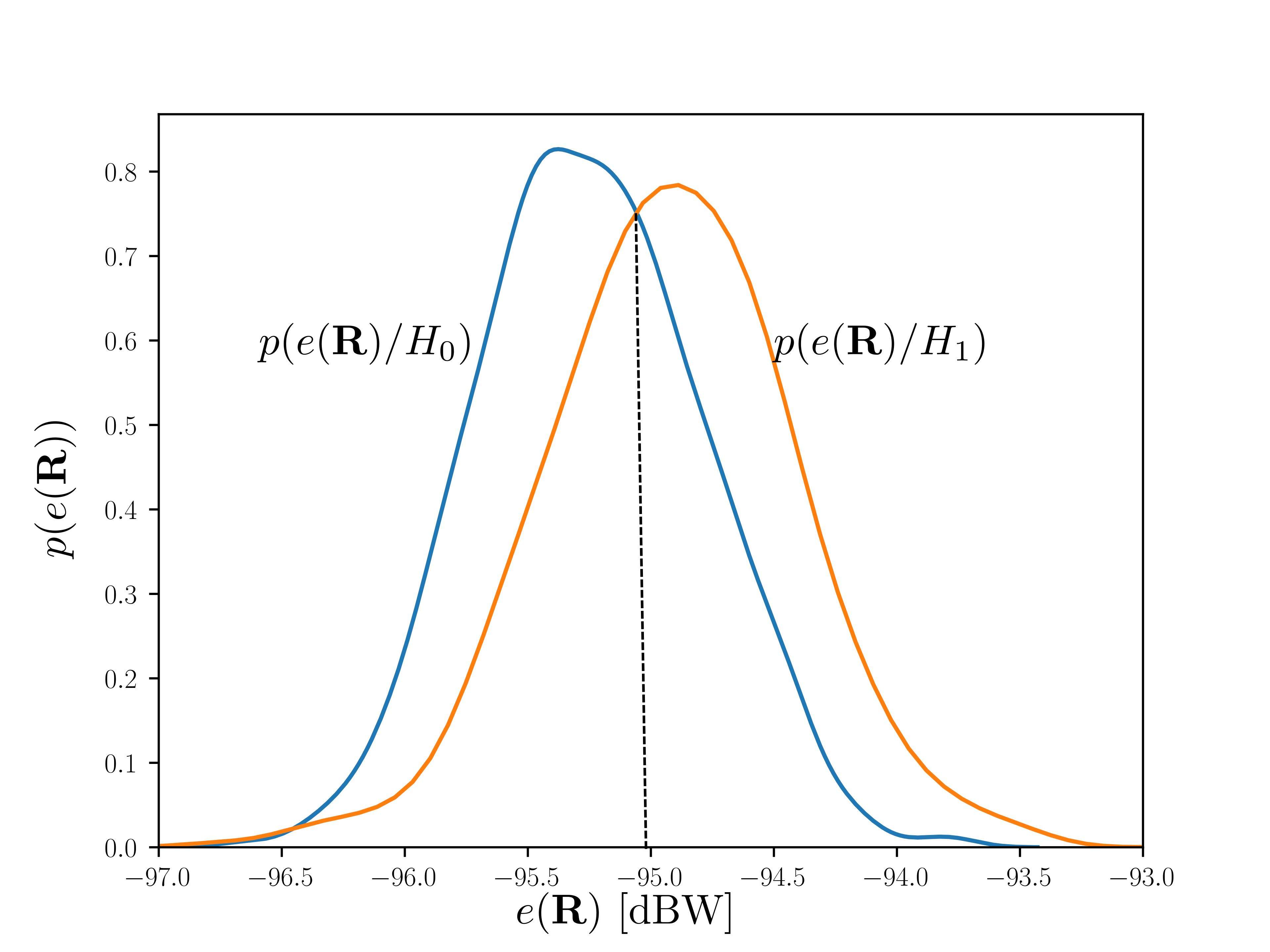}}
  \caption{Histogram of the delay-Doppler test-statistic under both hypothesis for moving target case.}
  \label{fig:his}
\end{figure}

\begin{figure*}
  \centering
  {\includegraphics[width=1.8\columnwidth]{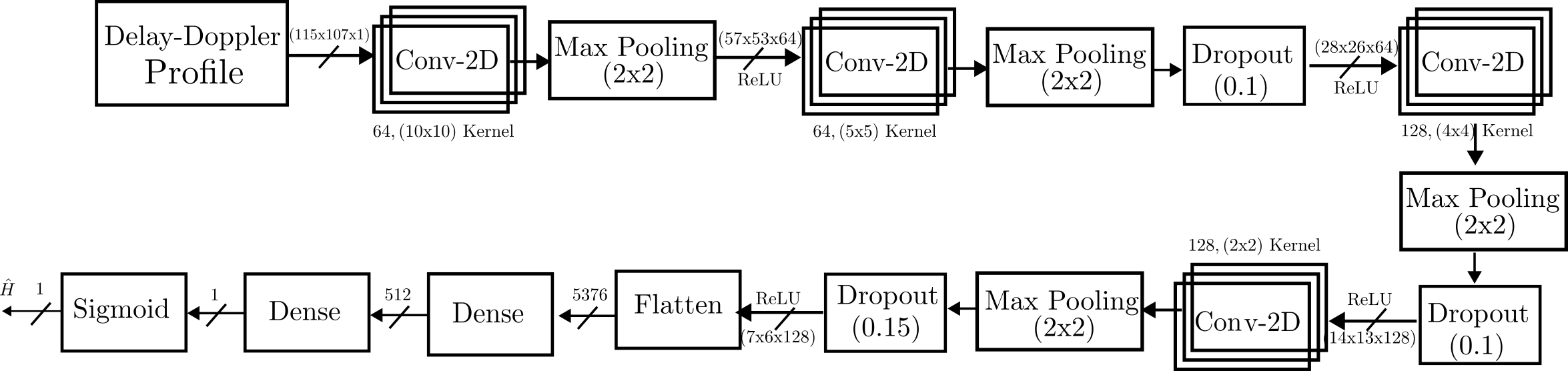}}
  \caption{AI architecture used for detecting the moving target using delay-Doppler profile.}
  \label{fig:arch}
\end{figure*}

\begin{figure}
  \centering
  \includegraphics[width=0.85\columnwidth]{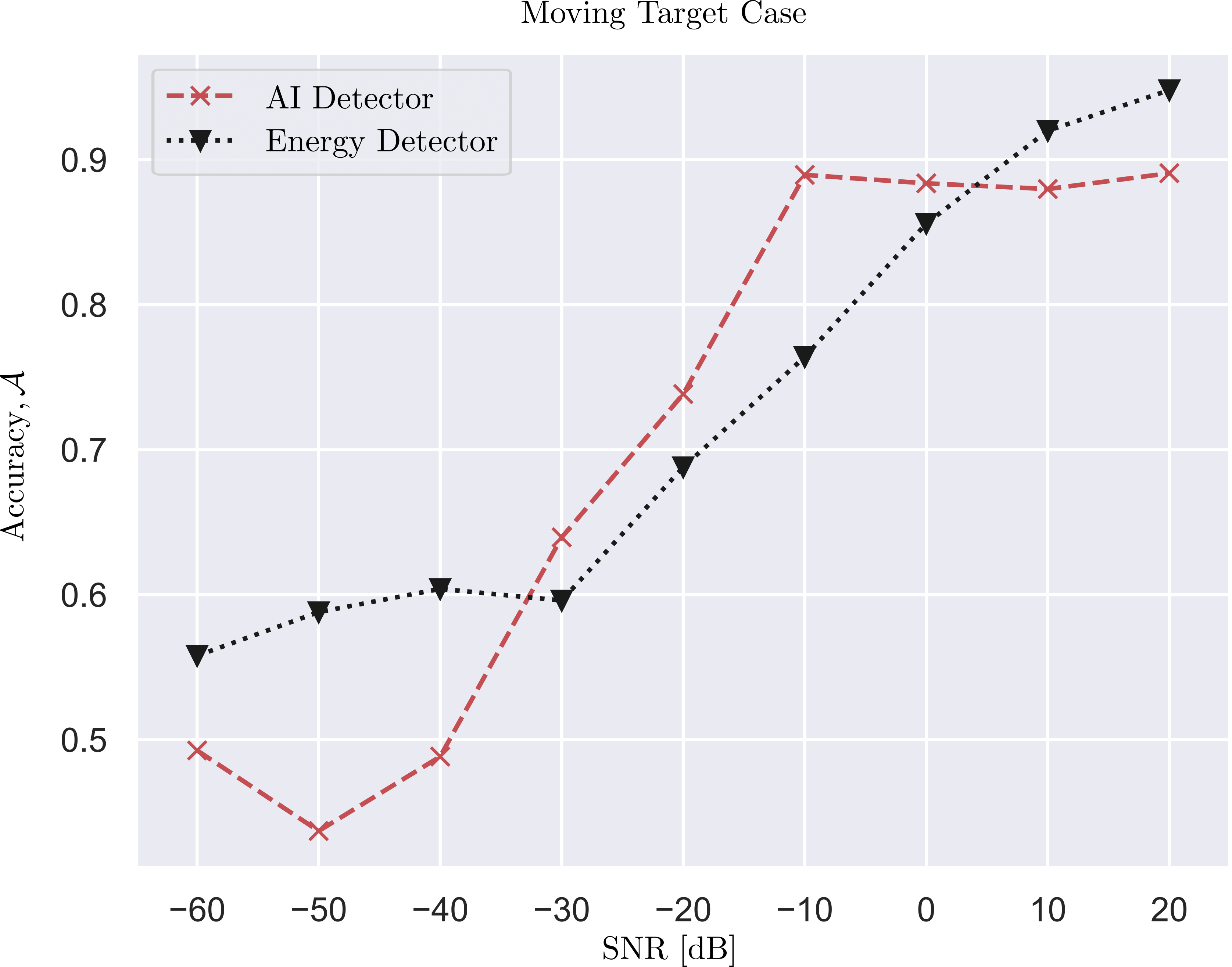}
  \caption{The performance of the AI detector and the baseline detector for moving target case.}
  \label{fig:los_ddp_performance}
\end{figure}

\begin{figure}
  \centering
  \includegraphics[width=0.85\columnwidth]{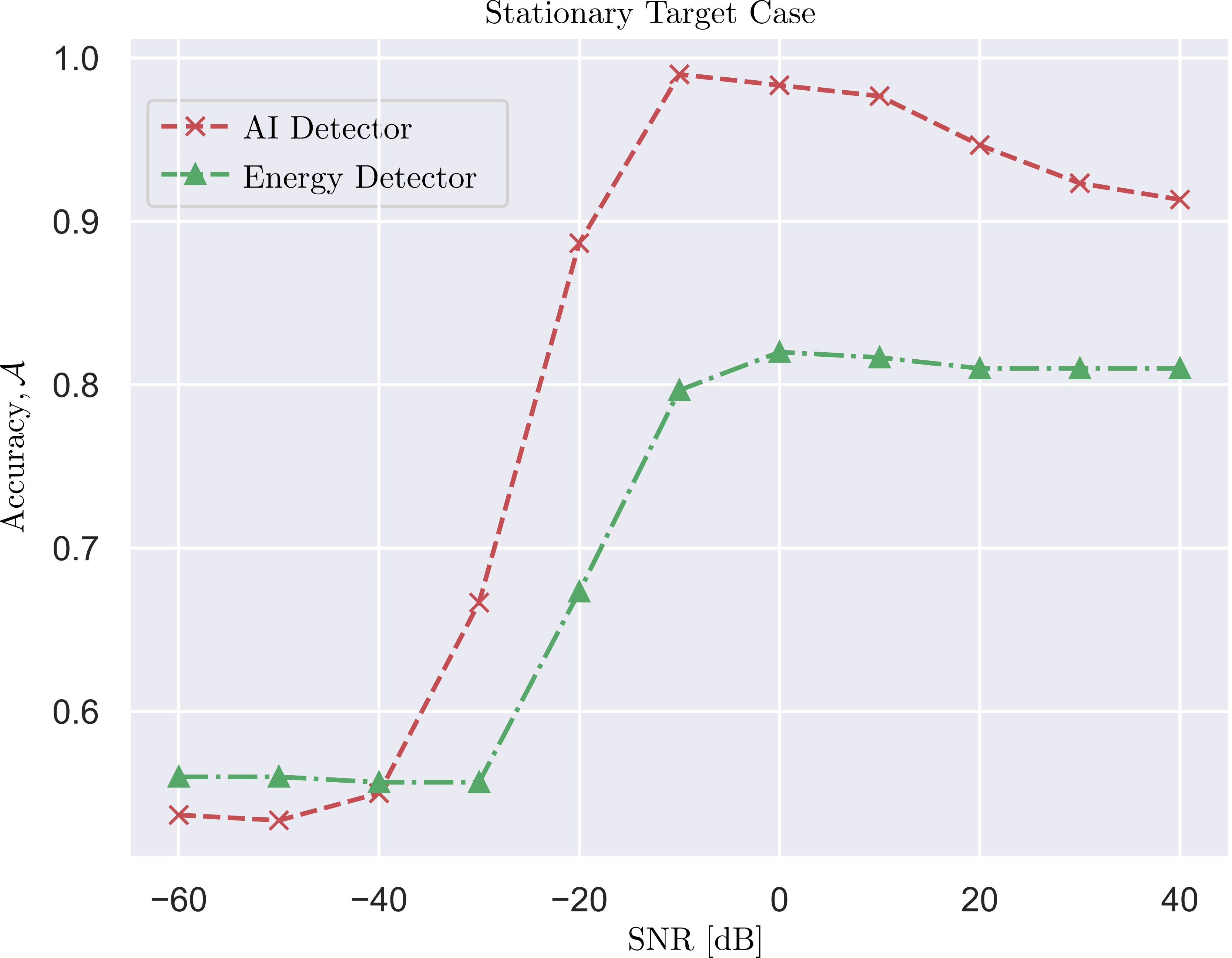}
  \caption{The performance of the AI detector and the baseline detector for stationary target case.}
  \label{fig:los_pdp_performance}
\end{figure}

\begin{figure}
  \centering
  \includegraphics[width=0.95\columnwidth]{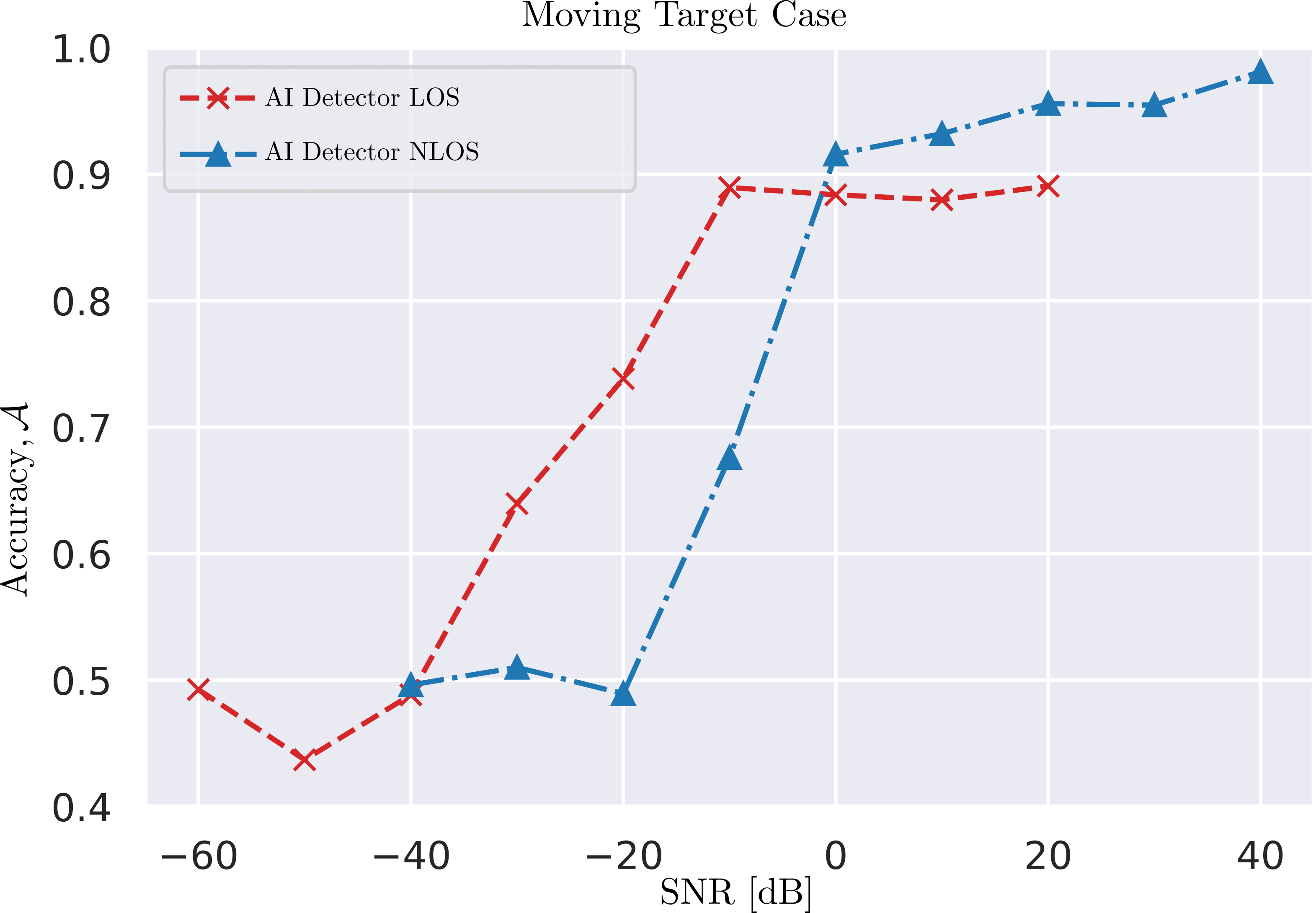}
  \caption{The performance of the AI detector for NLOS compared to LOS for moving target case.}
  \label{fig:nlos_ddp_performance}
\end{figure}

\begin{figure}
  \centering
  \includegraphics[width=0.8\columnwidth]{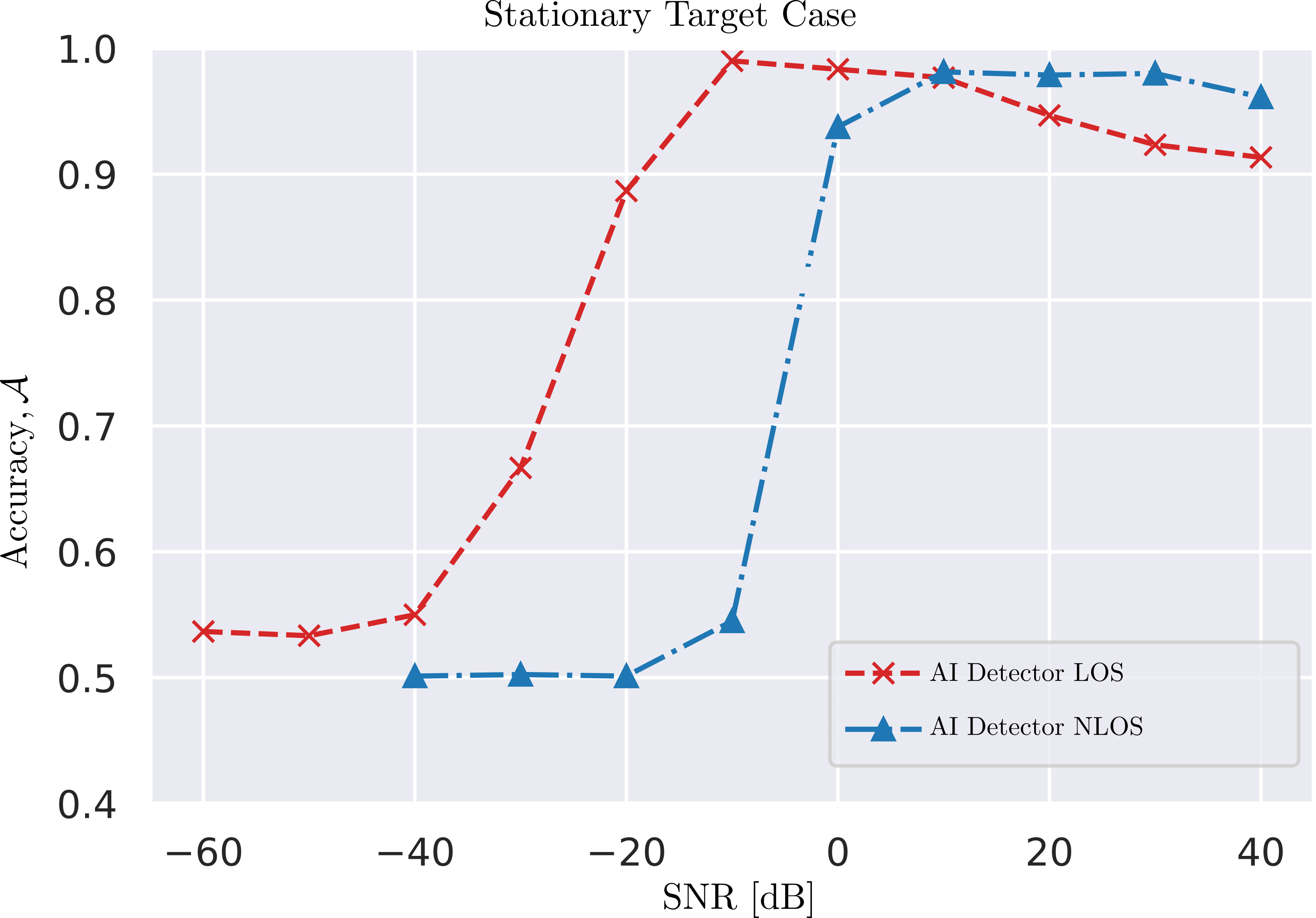}
  \caption{The performance of the AI detector for NLOS compared to LOS for stationary target case.}
  \label{fig:nlos_pdp_performance}
\end{figure}

We employ a \ac{CNN}-based AI architecture as shown in Fig.~\ref{fig:arch} similar to the proposed architecture in \cite{yajnanarayana2023eucnc}.  Analogous to typical image processing using AI, a $\mb{2D-CNN}$ pipeline is used to extract relevant features from a delay-Doppler frame for the moving target case. For the stationary target detection case, a similar architecture is employed using a $\mb{1D-CNN}$ pipeline.

For \ac{NLOS} setting, it is not straightforward to model  the received signal mathematically. We use the same AI pipeline architecture discussed above which uses \ac{DDP} and \ac{PDP} profiles from the received \ac{NLOS} signal towards target detection for moving and stationary target use cases respectively. 

\subsection{Baseline method}
Typical radar methods employ methods like  \ac{CFAR} detector for target detection, where for fixed false alarm probability ($P_{\MS{FA}}$), the detection probability ($P_{\MS{D}}$) is optimized. Since AI methods are configured to minimize the probability of error ($P_{\MS{e}}$) assuming the cost of error (Bayesian risk) to be same for both hypothesis, we compare the performance of the AI methods to a minimum $P_{\MS{e}}$ detector designed using energy thresholds. For an \ac{OFDM} receiver with both hypothesis having equal prior probability, the baseline detector can be expressed as

\begin{equation}
  \label{eq:ed}
p\left(e(\mbf{R})|H_1 \right) \mathop{\gtrless}_{H_0}^{H_1} p \left(e(\mbf{R})|H_0\right) ,
\end{equation}
where $p(\cdot)$ denotes the \ac{PDF} and $e( \cdot )$ denotes the test-statistic which is equal to the energy received by the receiver for the considered \ac{CPI}.An energy threshold, $\eta$  is used to decide on the hypothesis to minimize misclassification errors. The distributions and the thresholds are empirically calculated assuming a normal distribution with shifted mean values for $p\left(e(\mbf{R})|H_0 \right)$ and $p \left(e(\mbf{R})|H_1\right)$. For a moving target case, this is shown in Fig.~\ref{fig:his} and a threshold of $\eta=-95\,\mb{dBW}$ is used to separate the hypothesis.

\section{Simulations}
\label{sec:simulations}
\subsection{Training Procedure}
\label{ss:training}
We used an internal simulator to simulate the system model shown in Fig.~\ref{fig:system_model}. The transmit signal which is configured to the parameters defined in  Table~\ref{tab:params},  is  passed through a scattering point channel model with the  deployment environment shown in Fig.~\ref{fig:deployment} and Fig.~\ref{fig:deployment-nlos}. The received signal  is subjected \ac{DDP} or \ac{PDP} processing depending on whether the target is moving or stationary.  An AI agent is trained to understand the complex relationship between the DDP/PDP data and the hypothesis.

For LOS, we drop the target at $K$ random positions drawn randomly from XY-pane of the room having dimension $10\,\mb{m}\times 10\,\mb{m}$ (refer to Fig.~\ref{fig:deployment}) while for NLOS, the points are dropped between the walls (refer to Fig.~\ref{fig:deployment-nlos}). The resulting received signal is brought to delay-doppler plane and is used to build a labeled training set for alternate hypothesis ($H_{\MS{1}}$). Equal amount of null hypothesis records are similarly generated without the passive target in the room. The resulting labeled training set, $(\mathcal{D}_i,\mathcal{H}_i)\suchthat i=1,2,\ldots,2K$,  is used to train the AI agent shown in Fig.~\ref{fig:arch} for moving target case. Note that $\mathcal{D}_i$ and $\mathcal{H}_i\in\{H_0,H_1\}$  denote delay-doppler profile and  hypothesis respectively for the $i$-th record. For the stationary target case, we use the similar strategy to build the training set except  that here we use the \ac{PDP} instead of \ac{DDP} data. Therefore for stationary target detection case, the labeled training set, $(\mathcal{P}_i,\mathcal{H}_i)\suchthat i=1,2,\ldots,2K$, with $\mathcal{P}_i$ denoting the power delay profile for the $i$-th record is used to train the 1D-CNN-based AI pipeline discussed in Section~\ref{ss:arch}. Although we consider a single target in the simulations, the AI pipeline can be trained with data from a much larger input domain space having many targets of interest at various positions for multi-target sensing.

\subsection{Results}
The performance of the AI detector and the baseline detector for the target detection is  evaluated using accuracy score, $\mathcal{A}$:

\begin{multline}
  \mathcal{A} = 1-(p(\mathrm{target} | \mathrm{null})p(\mathrm{null}) \\
  + p(\mathrm{null} | \mathrm{target})p(\mathrm{target})),
\end{multline}
which is estimated empirically during testing. We generate $K=2000$ training samples for null ($H_0$) and alternate hypothesis ($H_1$) using the procedure described in the Section~\ref{ss:training}. A $70/30$ split is used to train and validate the corresponding AI pipelines designed for moving and stationary target cases. The performance of the trained AI detectors are compared with the energy detection based baseline detector. For testing the detectors, $150$ $H_1$ test samples are generated for both LOS and NLOS deployments. For LOS,  positions are randomly  drawn from  deployment in Fig.~\ref{fig:deployment} and for NLOS, the random positions are chosen in such a way to make sure they are NLOS to the receiver in the deployment of Fig.~\ref{fig:deployment-nlos}. For null hypothesis, $150$ $H_0$ samples are used for both LOS and NLOS respectively.
\subsubsection{LOS Channel}
\label{ss:los-results}
For the moving target case, the performance of the proposed AI detector and the baseline detector is as shown in Fig.~\ref{fig:los_ddp_performance}. The performance of the detection improves with SNR\footnote{SNR is defined as ratio of symbol energy to noise power spectral density ($E_s/N_0$), before the matched filter gain of $30\,\mb{dB}$.} for both AI and baseline detectors before plateauing out. However, the AI detector performance is $10\,\mb{dB}$ better in terms of SNR than the baseline detector at approximately $80$ percent detection accuracy. For the stationary case the performance is shown in the Fig.~\ref{fig:los_pdp_performance}, and here also we see  similar performance trend with AI detector outperforming the baseline detector.

\subsubsection{NLOS Channel}
\label{ss:nlos-results}
For the moving target case, the performance of the proposed AI detector for the \ac{NLOS} scenario is shown in Fig.~\ref{fig:nlos_ddp_performance} in comparison with the \ac{LOS} detector, while for the stationary target it is depicted in Fig.~\ref{fig:nlos_pdp_performance}. In both cases it can be seen that the AI detector is also able to detect the target in the \ac{NLOS} case and the curves behave similar. However, a $10-20\,\mb{dB}$ higher SNR is required compared to the LOS case. One possible reason could be due to the reflection loss incurred due to the walls resulting in reduced signal power at the receiver. Another interesting observation is the higher plateau for high SNR values in the case of a moving target for \ac{NLOS}, which could be due to the fact, that a moving scatterer can also result in two received paths at the receiver with different Doppler shifts, thereby improving detectability.

 \section{Discussion and Conclusions}
 The performance of the camera aided sensing systems with AI based detectors is superior in terms of  detection performance when compared to network based sensing, however they fare poorly when compared to the coverage, privacy and security. Network based sensing  addresses these challenges but it is challenging to achieve good detection performance even at high SNRs. We show that a shallow CNN based AI network trained using \ac{DDP} and \ac{PDP} can provide superior performance compared to baseline methods when used for moving and stationary target detection respectively. Results show that the performance of such an AI detector improves with SNR and outperforms the baseline detector's performance. The proposed method provides a gain of $10\,\mb{dB}$ compared to the baseline for $80$ percent detection rate. For NLOS channel, there is degradation of $10-20~\mb{dB}$ compared to LOS depending on the scenario. Based on the results, a human sized small object (having a typical $\mb{RCS}\approx 1\,\mb{m}^2$) in a cluttered room with several large objects (having typical $\mb{RCS}\gg 1\,\mb{m}^2$) can still be detected using a simple bi-static setup using proposed AI methods.

\bibliography{main}

\begin{thebibliography}{10}
\providecommand{\url}[1]{#1}
\csname url@samestyle\endcsname
\providecommand{\newblock}{\relax}
\providecommand{\bibinfo}[2]{#2}
\providecommand{\BIBentrySTDinterwordspacing}{\spaceskip=0pt\relax}
\providecommand{\BIBentryALTinterwordstretchfactor}{4}
\providecommand{\BIBentryALTinterwordspacing}{\spaceskip=\fontdimen2\font plus
\BIBentryALTinterwordstretchfactor\fontdimen3\font minus
  \fontdimen4\font\relax}
\providecommand{\BIBforeignlanguage}[2]{{%
\expandafter\ifx\csname l@#1\endcsname\relax
\typeout{** WARNING: IEEEtran.bst: No hyphenation pattern has been}%
\typeout{** loaded for the language `#1'. Using the pattern for}%
\typeout{** the default language instead.}%
\else
\language=\csname l@#1\endcsname
\fi
#2}}
\providecommand{\BIBdecl}{\relax}
\BIBdecl

\bibitem{yang20196g}
P.~Yang, Y.~Xiao, M.~Xiao, and S.~Li, ``{6G} wireless communications: Vision
  and potential techniques,'' \emph{IEEE network}, vol.~33, no.~4, pp. 70--75,
  2019.

\bibitem{wymeersch-2021-integ-commun}
\BIBentryALTinterwordspacing
H.~Wymeersch, D.~Shrestha \emph{et~al.}, ``Integration of communication and
  sensing in {6G}: a joint industrial and academic perspective,'' in \emph{2021
  IEEE 32nd Annual International Symposium on Personal, Indoor and Mobile Radio
  Communications (PIMRC)}, 9 2021, p. nil. [Online]. Available:
  \url{https://doi.org/10.1109/pimrc50174.2021.9569364}
\BIBentrySTDinterwordspacing

\bibitem{behravan-2022-introd}
\BIBentryALTinterwordspacing
A.~Behravan, R.~Baldemair \emph{et~al.}, ``Introducing sensing into future
  wireless communication systems,'' in \emph{2022 2nd IEEE International
  Symposium on Joint Communications \& Sensing (JC\&S)}, 3 2022, p. nil.
  [Online]. Available: \url{https://doi.org/10.1109/jcs54387.2022.9743513}
\BIBentrySTDinterwordspacing

\bibitem{zhang2021enabling}
J.~A. Zhang, M.~L. Rahman \emph{et~al.}, ``Enabling joint communication and
  radar sensing in mobile networks—a survey,'' \emph{IEEE Communications
  Surveys \& Tutorials}, vol.~24, no.~1, pp. 306--345, 2021.

\bibitem{de2021convergent}
C.~De~Lima, D.~Belot \emph{et~al.}, ``Convergent communication, sensing and
  localization in {6G} systems: An overview of technologies, opportunities and
  challenges,'' \emph{IEEE Access}, vol.~9, pp. 26\,902--26\,925, 2021.

\bibitem{levanon2000multi}
\BIBentryALTinterwordspacing
N.~Levanon, ``Multifrequency complementary phase-coded radar signal,''
  \emph{IEE Proceedings - Radar, Sonar and Navigation}, vol. 147, no.~6, p.
  276, 2000. [Online]. Available: \url{https://doi.org/10.1049/ip-rsn:20000734}
\BIBentrySTDinterwordspacing

\bibitem{sturm2011waveform}
C.~Sturm and W.~Wiesbeck, ``Waveform design and signal processing aspects for
  fusion of wireless communications and radar sensing,'' \emph{Proc. IEEE},
  vol.~99, no.~7, pp. 1236--1259, May 2011.

\bibitem{sen2010adapt-ofdm}
\BIBentryALTinterwordspacing
S.~Sen, ``Adaptive {OFDM} radar for target detection and tracking,''
  \emph{Washington Univ. in St. Louis}, 2010. [Online]. Available:
  \url{https://doi.org/10.7936/K7JQ0Z3C}
\BIBentrySTDinterwordspacing

\bibitem{yajnanarayana2023eucnc}
V.~Yajnanarayana and H.~Wymeersch, ``Multistatic sensing of passive targets
  using {6G} cellular infrastructure,'' in \emph{2023 Joint European Conference
  on Networks and Communications \& 6G Summit (EuCNC/6G Summit)}, 2023, pp.
  132--137.

\bibitem{saleh-1987-statis-model}
\BIBentryALTinterwordspacing
A.~Saleh and R.~Valenzuela, ``A statistical model for indoor multipath
  propagation,'' \emph{IEEE Journal on Selected Areas in Communications},
  vol.~5, no.~2, pp. 128--137, 1987. [Online]. Available:
  \url{https://doi.org/10.1109/jsac.1987.1146527}
\BIBentrySTDinterwordspacing

\bibitem{wilson-2010-radio-tomog}
\BIBentryALTinterwordspacing
J.~Wilson and N.~Patwari, ``Radio tomographic imaging with wireless networks,''
  \emph{IEEE Transactions on Mobile Computing}, vol.~9, no.~5, pp. 621--632,
  2010. [Online]. Available: \url{https://doi.org/10.1109/tmc.2009.174}
\BIBentrySTDinterwordspacing

\bibitem{chris2023eucnc}
C.~Mollen, G.~Fodor, R.~Baldemair, J.~Huschke, and J.~Vinogradova, ``Joint
  multistatic sensing of transmitter and target in {OFDM}-based {JCAS}
  system,'' in \emph{2023 Joint European Conference on Networks and
  Communications \& 6G Summit (EuCNC/6G Summit)}, 2023, pp. 144--149.

\end{thebibliography}
\bibliographystyle{IEEETran}
 
\end{document}